\documentclass[a4paper]{jpconf}
\usepackage{graphicx}
\usepackage{amssymb,amsmath}
\usepackage{color}

\begin{document}
\title{Where  do bosons  actually belong?}
%\shorttitle{Where bosons belong} %Insert here a short version of the title if it exceeds 70 characters

\author{A Marzuoli$^{\dag}$,  F A Raffa$^{\ddag}$ and M Rasetti$^{\sharp}$ }
%\shortauthor{A. Marzuoli \etal}

\address{$^{\dag}$ Universit$\grave{\textrm{a}}$ di Pavia, Dipartimento di Matematica $'$F. Casorati$\,'$,
and INFN Sezione di Pavia I-27100 Pavia, Italy}
  
\address{$^{\ddag}$ Politecnico di Torino, Dipartimento di Scienza Applicata e Tecnologia, I-10129 Torino, Italy}

 \address{$^{\sharp}$ Politecnico di Torino, Dipartimento di Scienza Applicata e Tecnologia, I-10129 Torino, and ISI Foundation, I-10133 Torino, Italy}

\ead{$^{\dag}$annalisa.marzuoli@pv.infn.it, $^{\ddag}$francesco.raffa@polito.it, $^{\sharp}$rasetti@isi.it}

\vspace{10pt}

\begin{flushright}
\footnotesize{PACS: 03.65.Fd \\ 
05.30.Jp \\
42.50.Ct }
\end{flushright}

\begin{abstract}
We explore a variety of reasons for considering $su(1,1)$ instead of the customary $h(1)$ as the natural unifying frame for characterizing boson systems. Resorting to the Lie--Hopf structure of these algebras, that shows how the Bose--Einstein statistics for identical bosons is correctly given in the $su(1,1)$ framework, we prove that quantization of Maxwell's equations leads to $su(1,1)$, relativistic covariance being naturally recognized as an internal symmetry of this dynamical algebra. Moreover $su(1,1)$ rather than $h(1)$ coordinates are associated to circularly polarized electromagnetic waves. As for interacting bosons, the $su(1,1)$ formulation of the Jaynes-Cummings model is discussed, showing its advantages over $h(1)$.
\end{abstract}

%\maketitle

\section{Introduction}
One of the pillars of quantum mechanics is the deep relation between what is referred to as 
$'$statistics$'$, systems of particles in nature obey either Fermi--Dirac or Bose--Einstein basic statistical counting rules, and the algebraic structure these imply. In such perspective, bosons have eventually come to be identified with the Weyl algebra, denoted by $h(1)$ and generated by creation and annihilation operators, operating over a Hilbert space of states assumed to be the Fock space. Such representation is tacitly assumed as natural and indisputable for free and interacting bosons, although a variety of foundational questions can be raised about the effective reliability of such choice. 
On the other hand, when bosons are photons, it is the relativistic covariance of the Maxwell equations that does matter. Describing photons in terms of the lowering and raising operators $a$ and $a^\dagger$, one sees that $h(1)$ does not admit the algebra of Lorentz transformations as internal symmetry. With this statement we mean the following: denoting by ${\mathfrak{U}}_{\rm SU(1,1)}$ the general unitary transformation of the group $SU(1,1)$, 
{\sl i.e.}, the unitary operator implementing a Lorentz 
transformation\footnote{There is a close relation of  $SU(1,1)$ with the Lorentz group: the special linear group $SL(2;\mathbb{R})$, sometimes referred to as Lorentz group by physicists, is indeed the double cover of the restricted Lorentz group $SO^{\uparrow}(1,3;\mathbb{R})$. Its identity component is $SO^{\uparrow}(1,2)$ $\cong$ $SL(2;\mathbb{R})/\mathbb{Z}_2$, whose algebra is $su(1,1)$. The latter is the algebra of the Lorentz group, {\sl cf.} 
\cite{GELF,BARG}.}, 
and by $\mathfrak{g}_{\rm h(1)}$ and $\mathfrak{g}_{\rm su(1,1)}$ generic Hermitian elements of Lie algebras $h(1)$ and $su(1,1)$, one can check that
\begin{equation}
{\mathfrak{U}}_{\rm SU(1,1)} \, \mathfrak{g}_{\rm su(1,1)} \, {\mathfrak{U}}_{\rm SU(1,1)}^{\dagger} \in su(1,1) \;\;,\;\; {\mathfrak{U}}_{\rm SU(1,1)} \, \mathfrak{g}_{\rm h(1)} \, {\mathfrak{U}}_{\rm SU(1,1)}^{\dagger} \notin h(1) \; . \label{TRASF}
\end{equation}
Eq. \eqref{TRASF} shows that, under any such unitary transformation, elements of $su(1,1)$ are mapped to themselves ({\sl i.e.}, the Lorentz group is an internal symmetry) while elements of $h(1)$ do not. 

These are the most serious obstructions to constructing theories characterized by the quadratic envelope of $h(1)$, {\sl i.e.}, the Weyl--Heisenberg (WH) algebra generated by 
$\{\mathbb{I}, a , a^{\dagger} , \hat{n} \} $, 
$\hat{n} = a^{\dagger} a$, as dynamical algebra that exhibits both correct statistics and relativistic covariance.
Here we deal with the delicate question: is there a different way to represent bosons avoiding these difficulties and providing a  unified  dynamical framework?  We argue that the representation of bosons should be grounded in the algebra $su(1,1)$, the semi--simple algebra of the non compact Lie group $SU(1,1)$ generated by $\{ K_+,K_-, K_3\}$
or $\{ K_1,K_2, K_3\}$ with 
$K_+ = K_1+\imath K_2$ and $K_-=K_1-\imath K_2$. 
The reasons for such claim are manifold and we touch on them discussing the emerging structure 
{\sl{vs.}} the traditional one. The paper is organized following the various facets of our argument. 1) Considering the quantum statistics side of the problem in the frame of the associated Hopf algebra, a collection of bosons can be consistently considered a bosonic object, satisfying the basic requirements of Bose--Einstein statistics, only within the coalgebra of $su(1,1)$; $h(1)$ leads instead to Maxwell--Boltzmann statistics. This point, which has been raised previously by one of the authors, is discussed in  section 3 for completeness. 2) When performing the quantization of the normal modes of the electromagnetic field, based {\sl{ab initio}} on Maxwell's equations, the most general algebraic frame for quantization is $su(1,1)$ rather than $h(1)$, since the canonical structure of the classical equations of motion allows for more freedom than that of a collection of (infinitely many) harmonic oscillators. Moreover, the choice of $su(1,1)$ permits quite naturally the implementation of Lorentz covariance, which $h(1)$ does not. Finally, the group generated by the Poisson brackets of the classical normal mode--field polar components in the plane orthogonal to the propagation direction is $SO(1,2)$, the proper Lorentz group in (2+1) space--time dimensions; canonical quantization  requires to go through the isomorphism $so(1,2)  \sim su(1,1)$. 3) $su(1,1)$ realizations of the quantum harmonic oscillator retain most of the physical properties of the realization in WH. 4) When the boson ({\sl e.g.}, photon) is not free but interacts with matter, say a two--level atom as in the Jaynes--Cummings model, its description in terms of $su(1,1)$ ladder operators does not alter the cycles of atomic collapse and revival, integrable regardless of the initial state of the system.
%%%%%%%%%%%%%%%%%%%%%%%%%%%%%%%%%%%%%%%%%%%%%%%%%%%%%%
%%%%%%%%%%%%%%%%%%%%%%%%%%%%%%%%%%%%%%%%%%%%%%%%%%%%%%%
\section{Algebras and representations}
%%%%%%%%%%%%%%%%%%%%%%%%%%%%%%%%%%%%%%%%%%%%%%%%%%%%%%
%%%%%%%%%%%%%%%%%%%%%%%%%%%%%%%%%%%%%%%%%%%%%%%%%%%%%%%
The basic tool for a fully consistent characterization of bosons is the Lie--Hopf structure associated to a Lie algebra ${\mathcal{L}}$ (here $h(1)$ or $su(1,1)$) 
\cite{BARG,FUCHS,BOOKS}. The motivation why the Hopf algebra is necessary in describing statistics is that, whenever a physical object is characterized by a dynamical algebra, the requirement that a collection of objects of the same nature belongs to the same category reflects mathematically in the requirement that the dynamical algebra of the many-particles system is the coalgebra of the single-object dynamical algebra
\footnote{We mention that Hopf algebras have been used in the context of many-boson systems in \cite{SOLO} where, however, their physical meaning is different from the one of the present work.}. The composition of angular momenta is a good example of this point. Thus the rationale for the introduction of such enveloping algebras  relies on the necessity of switching from the non--associative structure of ${\mathcal{L}}$ to an associative algebra together with its coalgebra, which encode the essential properties of $\mathcal{L}$.
Here the relevant map is the coproduct $\Delta(X)$ $=$ $\displaystyle X \otimes \mathbb{I} + \mathbb{I} \otimes X$ $=$ $X_1 + X_2$, where $X$ $\in$ $\cal{L}$, $\mathbb{I}$ is the identity operator and $X_j$ denotes operator $X$ in particle-$j$ Hilbert space, $j=1,2$.

Basic difficulties encountered with the Weyl algebra $h(1)$ --mentioned in passing in  the introduction--
are faced when moving from the single--particle to the many--particle picture: 
i) $h(1)$, as defined by the conventional canonical commutation relations
$\bigl [ a , a^{\dagger} \bigr ] = 1$, is not the enveloping algebra of some Lie algebra. The WH  algebra, enveloping algebra of $h(1)$ thought of as its ideal, is neither semi--simple nor simple; ii) $h(1)$ has no finite dimensional representations, as one can check by the contradiction encountered taking the trace of the basic commutation relation; iii) there is no Hopf algebra structure compatible with this commutation relation 
\cite{BORO}: also the counit map applied to both sides of it leads to an absurd, as  $\varepsilon (1) = 1$, while the form of primitive coproduct standard in the case of Lie algebras, $\Delta (a) = a \otimes {\mathbb{I}} + {\mathbb{I}} \otimes a$, does not fulfill the requirement of co--commutativity and poses a further obstruction to a consistent many--particle structure.

As for $su(1,1)$ representations, recall that only the positive, discrete series irreducible unitary representations ${\cal D}^{(+)}_{\kappa}$, indexed by $\kappa$ $=$ $\frac{1}{2}, 1, \frac{3}{2}, 
2 \ldots$, and the anomalous irrep. ${\cal{D}}_{\frac{1}{4}}$ $(\kappa = \frac{1}{4})$ are spanned by the complete orthonormal set $\{ |\kappa ,n \rangle \, |\, n \in {\mathbb{N}} \}$ of eigenstates of $K_3$. Each $|\kappa ,n \rangle$ can be identified for any given $\kappa$ with the eigenstate $|n\rangle$ of $\hat{n}$ in Fock space ${\mathfrak{F}}$ $=$ ${\rm{span}} \{ | n \rangle \, | \, n \in {\mathbb{N}} \}$, with highest weight vector $|0\rangle$ annihilated by the lowering operator, $K_- |0\rangle \doteq 0$. We shall consider ${\cal D}^{(+)}_{\kappa}$ in the Holstein--Primakoff representation \cite{HP} for which,  over $\mathfrak{F}$,
\begin{equation}
K_3 = \hat{n}+\kappa \;\;,\;\; K_- = (K_+)^{\dag} = \sqrt{\hat{n}+2\kappa}\,a \; . \label{HP}
\end{equation}
$\hbar = 1$ is used throughout. 
%%%%%%%%%%%%%%%%%%%%%%%%%%%%%%%%%%%%%%%%%%%%%%%%%%%%%%
%%%%%%%%%%%%%%%%%%%%%%%%%%%%%%%%%%%%%%%%%%%%%%%%%%%%%%%
\section{Boson statistics and algebras}
%%%%%%%%%%%%%%%%%%%%%%%%%%%%%%%%%%%%%%%%%%%%%%%%%%%%%%%%%%
%%%%%%%%%%%%%%%%%%%%%%%%%%%%%%%%%%%%%%%%%%%%%%%%%%%%%%%%%
Following the seminal discussion in \cite{ROMAN}, we analyze the statistical counting behavior of a system of $n \geq 0$ (fixed) identical particles, where each particle can be in any of $m$ possible states.
In particular, we want to evaluate the distribution function $\displaystyle {\cal P}_n(k_1, \cdots , k_m)$, {\sl i.e.}, the probability of finding $k_j$ particles in mode $j$, $j = 1, \dots , m$, with the constraint $ k_1 + \cdots + k_m$ $=$ $n$. To this aim,  in the space state $\mathfrak{H}$ $=$ $\mathfrak{H}_1^{\otimes m}$, where $\mathfrak{H}_1$ is the single-mode Hilbert space, we define the $m$-mode vacuum, 
\begin{equation}
|0\rangle_m \doteq {\underbrace{|0\rangle \otimes \cdots \otimes |0\rangle}_{m \; {\rm times}}}
\end{equation}
and the extended coproduct $\Delta_m (\mathfrak{a}^\dagger) \equiv  \mathfrak{a}_1^\dagger + \cdots + \mathfrak{a}_m^\dagger$, 
\begin{equation}
\Delta_m (\mathfrak{a}^\dagger) = \sum_{j=1}^m {\mathbb{I}}^{\otimes (j-1)} \otimes {\mathfrak{a}}^{\dagger} \otimes {\mathbb{I}}^{\otimes (m-j)} \; , \nonumber
%\label{COmMODE}
\end{equation}
where $\mathfrak{a}^\dagger$ denotes the single--mode creation operator of $\cal{L}$ and $\displaystyle \mathfrak{a}_j^\dagger$ acts on the $j$-th slot of $|0\rangle_m$, $j$ $=$ $1, \dots , m$. We construct then the $m$-mode state $|\Phi\rangle$ $=$ $(\Delta_m (\mathfrak{a}^\dagger))^n |0\rangle_m$, so that $\displaystyle {\cal P}_n(k_1, \cdots , k_m)$ follows from the explicit expression of $|\Phi\rangle$ in terms of product states. Since for $h(1)$ and $su(1,1)$ $\displaystyle [\mathfrak{a}_i^\dagger,\mathfrak{a}_j^\dagger]$ $=$ 0 for $i \neq j$, map $\bigl ( \Delta_m (\mathfrak{a}^\dagger) \bigr )^n$ for both algebras proves to be
\begin{equation}
{\sum_{k_j}}' n! \, \prod_{j = 1}^m \left ( k_j ! \right )^{- \frac{1}{2}} \, {\mathfrak{a}^\dagger}^{k_1} \otimes \cdots \otimes {\mathfrak{a}^\dagger}^{k_m} \; ,\label{COPMAPn}
\end{equation}
where $\sum_{k_j}'$ means sum over the $m$ numbers $k_j \geq 0$ with $ k_1 + \cdots + k_m$ $=$ $n$. For $h(1)$ one has $\mathfrak{a}^\dagger$ $\equiv$ $a^\dagger$ in \eqref{COPMAPn} (with $\displaystyle {a^\dagger}^k |0\rangle$ $=$ $\sqrt{k!}|k\rangle$\,) and the normalized state $\displaystyle |\Phi\rangle$ reads
$$
|\Phi\rangle = \left ( m^n \, n! \right )^{- \frac{1}{2}} {\sum_{k_j}}' \, 
n! \, \left ( \prod_{j = 1}^m k_j ! \right )^{- \frac{1}{2}} \, |k_1, \dots, k_m\rangle \; ,
$$
so that ${\cal P}_n(k_1, \dots , k_m)$ is the multinomial probability distribution with probabilities $p_1$ $=$ $\dots$ $=$ $p_m$ $=$ $1/m$
\begin{equation}
{\cal P}_n(k_1, \dots , k_m) = m^{-n} \, n! \, \, \prod_{j = 1}^m 
\left ( k_j ! \right )^{- 1} \; . \label{PROBh1}
\end{equation}
Adopting instead in \eqref{COPMAPn} the fundamental representation $\kappa$ $=$ $1/2$ of 
$su(1,1)$, which gives $\mathfrak{a}^\dagger$ $\equiv$ $K_+$ $=$ $a^\dagger \sqrt{\hat{n}+1}$ $\displaystyle (K_+^k |0\rangle$ $=$ $k ! |k\rangle)$, the normalized state $\displaystyle |\Phi\rangle$ is 
$$
|\Phi\rangle = \left ( (m-1)! \, \prod_{j=1}^{m-1} (n+j)^{- 1} \right )^{\frac{1}{2}}  {\sum_{k_j}}'  \,|k_1 , \dots , k_m\rangle 
$$
and the corresponding probability distribution is uniform
\begin{equation}
{\cal P}_n(k_1 , \dots , k_m) = (m-1)! \, \prod_{j=1}^{m-1} (n+j)^{-1} \; . \label{PROBsu11}
\end{equation}
Comparison of \eqref{PROBh1} and \eqref{PROBsu11} shows that while $h(1)$ is naturally associated with the classical statistics of $n$ objects distributed in $m$ identical slots with equal probability $1/m$, with  $su(1,1)$ one obtains a uniform probability distribution for all $m$ accessible states, which corresponds physically to the case of $n$ identical bosons. Therefore, in terms of a pure statistical counting, $h(1)$ induces the classical Boltzmann--Maxwell probability distribution while the Bose--Einstein statistics is related to $su(1,1)$. Note that this does not change the usual B.E. distribution for non--zero  temperature, that requires only that the occupation number is unconstrained, as reported in \cite{CR}. Note as well that in this section, on the one side, we generalize the algebraic approach detailed in \cite{CRV-CR} for $m = 2$, while, on the other side, we refer to the statistics of coproduct states, unlike \cite{CR} where generalized coherent states of $su(1,1)$ were considered.
%%%%%%%%%%%%%%%%%%%%%%%%%%%%%%%%%%%%%%%%%%%%%%%%%%%%%%
%%%%%%%%%%%%%%%%%%%%%%%%%%%%%%%%%%%%%%%%%%%%%%%%%%%%%%%
\section{Maxwell's equations and algebras}
%%%%%%%%%%%%%%%%%%%%%%%%%%%%%%%%%%%%%%%%%%%%%%%%%%%%%%%%%%
%%%%%%%%%%%%%%%%%%%%%%%%%%%%%%%%%%%%%%%%%%%%%%%%%%%%%%%
Let us write  Maxwell's equations in the vacuum in the Riemann--Silberstein complex formalism \cite{RIESIL} 

\begin{equation} 
\mathbf{\nabla} \cdot \mathbf{G} = \frac{\rho}{\varepsilon_0} \quad,\quad \mathbf{\nabla} \times \mathbf{G} = \frac{\imath}{c}  \frac{\partial \mathbf{G}}{\partial t} + \imath \sqrt{\frac{\mu_0}{\varepsilon_0}} \mathbf{j} \label{MAXW} \; ,
\end{equation} 
where $\mathbf{G} \doteq \mathbf{E} + \imath c \mathbf{B}$, $\rho$ and $\mathbf{j}$ are the charge and current densities, $\varepsilon_0$ and $\mu_0$ are the vacuum electric permettivity and magnetic permeability, respectively, and $c^2$ $=$ $(\varepsilon_0 \mu_0)^{-1}\;$. The mode expansion of $\mathbf{G}$ is
\begin{equation}\label{Gi}
\mathbf{G}(\mathbf{x}) = \frac{1}{\sqrt{V}} \sum_{\mathbf{k}} e^{\imath \mathbf{k \cdot x}} \left(a_\mathbf{k} \mathbf{e}^{(1)}_\mathbf{k} + b_\mathbf{k} \mathbf{e}^{(2)}_\mathbf{k} + c_\mathbf{k} \mathbf{e}^{(3)}_\mathbf{k} \right) \; ,
\end{equation}  
where $V$ denotes the volume of the system endowed with periodic boundary conditions, $\mathbf{k}$ is the wave vector, $\mathbf{e}^{(\alpha)}_\mathbf{k}$, $\alpha = 1,2,3$, are three orthonormal vectors, with $\mathbf{e}^{(3)}_\mathbf{k}$ $=$ $\mathbf{k}/k$, $k = |\mathbf{k}|$, and $a_\mathbf{k}$, $b_\mathbf{k}$, $c_\mathbf{k}$ are complex scalars. In the absence of charges and currents, $c_\mathbf{k}$ $\equiv$ 0 and the electromagnetic field is fully described by $a_\mathbf{k}$ and $b_\mathbf{k}$. Splitting $a_\mathbf{k}$ and $b_\mathbf{k}$ in their real and imaginary parts, $a_\mathbf{k}$ $\doteq$ $ (p_\mathbf{k}^{(1)} + \imath p_\mathbf{k}^{(2)})$, $b_\mathbf{k}$ $\doteq$ $k c (q_\mathbf{k}^{(1)} + \imath q_\mathbf{k}^{(2)})$, the equations of motion read, for each ${\bf{k}}$,  
\begin{equation}
\dot{p}_\mathbf{k}^{( \alpha )} = - (c k)^2 q_\mathbf{k}^{( \alpha )} \; ,\; \dot{q}_\mathbf{k}^{( \alpha )} =  p_\mathbf{k}^{( \alpha )} \; ; \; \alpha = 1,2 \; . \label{MAXreim}
\end{equation}
Clearly, Eqs. \eqref{MAXreim} can be thought of as obtained in canonical form from a Hamiltonian $H$, which is the sum of infinitely many Hamiltonians of  uncoupled harmonic oscillators of two types: 
$$ 
H = \frac{1}{2} \sum_{\mathbf{k}} \bigl ( \bigl | a_\mathbf{k} \bigr |^2 + \bigl | b_\mathbf{k} \bigr |^2 \bigr ) =\sum_{\mathbf{k}}\,  \sum_{\alpha = 1,2} \frac{1}{2} \Bigl ( {p_\mathbf{k}^{( \alpha )}}^2 + \omega_k^2 {q_\mathbf{k}^{( \alpha )}}^2 \Bigr )
$$ 
with $\omega_k \doteq c k$. 
Usually, in this perspective, quantization simply proceeds along the lines of conventional quantum theory of radiation \cite{SAKU}: $a_\mathbf{k}$, $\bar{a}_\mathbf{k}$ and $b_\mathbf{k}$, $\bar{b}_\mathbf{k}$ are identified with the creation and annihilation operators ${\mathfrak{a}}_{\alpha}^{\mathbf{k}}$, ${\mathfrak{a}}_{\alpha}^{\mathbf{k} \, \dagger}$, $\alpha = 1,2$, satisfying the standard commutation relations of two independent $h(1)$ for each ${\bf{k}}$. With $\hat{n}_{\alpha}^{\mathbf{k}}$ $\doteq $ ${\mathfrak{a}}_{\alpha}^{\mathbf{k} \, \dagger} {\mathfrak{a}}_{\alpha}^{\mathbf{k}}$, the occupation number operator of mode $(\mathbf{k},\alpha )$, the quantum Hamiltonian has the standard  form 
$$
H \,=\, \sum_{\mathbf{k}} \sum_{\alpha =1,2} H_\mathbf{k}^{( \alpha )},\;\;\, H_\mathbf{k}^{( \alpha )} =
\omega_k \bigl ( \hat{n}_\alpha^{(\mathbf{k})} + \tfrac{1}{2} \bigr ).
$$
Note that the vacuum energy factor $\tfrac{1}{2}$ implies the divergence whereby the theory needs to be renormalized. The dynamical algebra $\mathfrak{A}$ in this case is given by the direct sum over the modes ${\mathbf{k}}$: $\mathfrak{A} = \oplus_{\mathbf{k}}^{} \mathfrak{A}_\mathbf{k}$, 
$\mathfrak{A}_\mathbf{k} = h(1)_\mathbf{k} \oplus h(1)_\mathbf{k}$.  We argue that the above quantization procedure is by no means unique. One can indeed define a new formal bracket, $\{\bullet,\bullet\}$, such that the equations of motion of the two independent degrees of freedom of mode ${\bf{k}}$ can be written as 
$$
{\dot{p}}_\mathbf{k}^{( \alpha )} = \{ p_\mathbf{k}^{( \alpha )} , H_\mathbf{k}^{( \alpha )} \} \; ; \; \;{\dot{q}}_\mathbf{k}^{( \alpha )} = \{ q_\mathbf{k}^{( \alpha )}, H_\mathbf{k}^{( \alpha )} \} \; , \; \alpha = 1,2 \; ,
$$
provided the following equations are verified, $\forall \, {\bf{k}}$,  $\alpha$ 
$$
\{ p_\mathbf{k}^{( \alpha )} , H_\mathbf{k}^{( \alpha )} \} = - \frac{\partial H_\mathbf{k}^{( \alpha )}}{\partial q_\mathbf{k}^{( \alpha )}} \; ; \; \{ q_\mathbf{k}^{( \alpha )} , H_\mathbf{k}^{( \alpha )} \} = \frac{\partial H_\mathbf{k}^{( \alpha )}}{\partial p_\mathbf{k}^{( \alpha )}} \; .
$$
The requirement holding for the customary Poisson brackets ($PB$) that $\displaystyle{ \bigl \{ q_{\mathbf{k}}^{( \alpha )} , p_{\mathbf{k}'}^{( \beta )} \bigr \}_{PB} = \delta_{\alpha , \beta} \, \delta_{{\mathbf{k}},{\mathbf{k}'}}}$ can be dropped, as for mode ${\bf{k}}$ neither position nor momentum conjugate variables need to be defined {\sl{a priori}}. The new bracket however still returns the desired equations of motion if one requires that    
$$
\{ q_{\mathbf{k}}^{( \alpha )} , p_{\mathbf{k}'}^{( \beta )} \} = \delta_{\alpha , \beta} \, \delta_{{\mathbf{k}},{\mathbf{k}'}} \,   \mathcal{I}_{\mathbf{k}}^{( \alpha )} \; ,
$$
where $\mathcal{I}_\mathbf{k}^{( \alpha )}$ $=$ $\mathcal{I}_\mathbf{k}^{( \alpha )} \bigl ( q_{\mathbf{k}}^{( \alpha )} , p_{\mathbf{k}}^{( \alpha )} \bigr )$, are constants of the motion, i.e., 
$\displaystyle{ \bigl \{ \mathcal{I}_{\mathbf{k}}^{( \alpha )} , H_{\mathbf{k}'}^{( \beta )} \bigr \} = 0}$, $\forall \, {\bf{k}} , {\bf{k}}' , \alpha , \beta$.
A minimal choice to satisfy such requirement consists in choosing  $\mathcal{I}_\mathbf{k}^{( \alpha )}$ $=$ $\lambda^{( \alpha)} H_\mathbf{k}^{( \alpha )}$, where $\lambda^{( \alpha )}$ are $c$-numbers independent on $t$ and on $q_{\mathbf{k}}^{( \alpha )}$, $p_{\mathbf{k}}^{( \alpha )}$. 
In this case the dynamical algebra turns out to be  $\mathfrak{A}$ $=$ $\displaystyle{\oplus_{\mathbf{k}}^{}\; \mathfrak{A}_\mathbf{k}}$, where we have now just that, $\forall \,\mathbf{k}$, 
$$
\mathfrak{A}_\mathbf{k} \,=\,\bigoplus_{\alpha = 1,2} su(1,1)_{\mathbf{k}}^{( \alpha )} \,.
$$ 
Indeed, upon setting 
\begin{align}
q_{\mathbf{k}}^{( \alpha )}& \equiv \tfrac{1}{2} (K_{+}^{( \alpha )} + K_{-}^{( \alpha )}) \; , \nonumber\\
p_{\mathbf{k}}^{( \alpha )}& \equiv \tfrac{\imath }{2} (K_+^{( \alpha )} - K_-^{( \alpha )}) \; , \nonumber\\ 
H_{\mathbf{k}}^{( \alpha )}& =\omega_k  (K_3^{( \alpha )} - \kappa + 1/2) \; , \nonumber
\end{align}
so that 
$$
H = \sum_{\bf{k}} \sum_{\alpha = 1,2} H_{\mathbf{k}}^{( \alpha )}\, ,
$$
operators $K_\lambda^{( \alpha )}$, $\lambda \in \{+,- , 3\}$, generate $su(1,1)_{\mathbf{k}}^{( \alpha )}$. Note that for $\kappa = 1/2$ the energy spectrum is identical to that of a conventional harmonic oscillator and the theory is automatically regularized.
%%%%%%%%%%%%%%%%%%%

Relativistic invariance of the Maxwell equations is dealt with recalling that a Lorentz transformation  for the mode operators 
${\mathfrak{a}}_{\alpha}^{\mathbf{k} \, \dagger}$,  ${\mathfrak{a}}_{\alpha}^{\mathbf{k}}$, 
with  $\gamma = \sqrt{1 - v^2/c^2}\,$ and
$$
\mathbf{k}' = {\mathbf{k}} + \frac{\gamma^2 - 1}{\gamma} \, \omega_{\mathbf{k}} \left[ 1 + {\sqrt{\frac{\gamma - 1}{\gamma + 1}}} \, \cos ( \bf{k} \cdot \bf{v} ) \right] \frac{\bf{v}}{v^2} \; ,
$$
reads 
$$ \bigl | \, {{\mathfrak{a}_1^{\mathbf{k'}}}} \, , {{\mathfrak{a}_2^{\mathbf{k'}}}} \bigr \rangle \;=\; \left[\mathbb{M}\right]\, \bigl | \,  {\mathfrak{a}_1^{\mathbf{k}}} \; , {\mathfrak{a}_2^{\mathbf{k}}} \bigr \rangle \,.
$$ 
Here primed quantities refer to a system ${\cal{S}}^\prime$ which moves with respect to system ${\cal{S}}$ with velocity $\mathbf{v} = v \mathbf{e}_z$ and the matrix $[\mathbb{M}]$ is 
\begin{equation}
\left[\mathbb{M}\right]  = 
\left[ \begin{array}{cc}
\gamma
&
\imath \sqrt{\gamma^2 - 1}
\\
- \imath \sqrt{\gamma^2 - 1}
&
\gamma
\end{array} \right] \; . \label{MLOR}
\end{equation} 
Note that matrix \eqref {MLOR} is Hermitian, orthogonal but not unitary, and, with $\gamma \doteq \cosh(\frac{1}{2}\, \vartheta)$, $\sqrt{\gamma^2 - 1}$ $=$ $\sinh(\frac{1}{2}\, \vartheta)$, it describes a hyperbolic rotation of $\vartheta$ around the unit vector ${\bf{e}}_x$ \cite{PERE1} and is an element of $SU(1,1)$. In particular $\left[\mathbb{M}\right]$ is obtained by the exponential map $\displaystyle \exp{[\imath \vartheta \, (K_+ + K_-)/2]}$ of generators of the fundamental representation of $su(1,1)$, which is isomorphic to the Lie algebra $so(1,2)$ of the Lorentz group ${\mathfrak{L}}$
\footnote{The algebra $su(1,1)$ has been introduced in a quantum optics context by Yurke et al. \cite{YURKE}, though not as dynamical algebra of photons but as the algebra of the $'$transfer functions$'$ of a number of relevant optical devices.}. By resorting to the adjoint map ${\mathfrak{L}}$ $\rightarrow$ Aut($so(1,2)$) the realization of $su(1,1)$ as an inner automorphism of the dynamical group is accomplished. In other words, covariance under the Lorentz group of transformations ${\mathfrak{L}}: {\cal{S}} \mapsto {\cal{S}}^\prime$ is naturally realized in the $su(1,1)$ framework, see Eq. \eqref{TRASF}. Whereas for ${\mathfrak{L}}$ to be realized as an inner automorphism of $h(1)$, the corresponding $su(1,1)$ should be contracted ($\kappa$ or $c$ $\to \infty$), which however would turn ${\mathfrak{L}}$ into the classical Galilei group \cite{INOWIG}.

Note finally that, with condition $c_\mathbf{k}$ $=0$ in {Eq. \eqref{Gi}}, the complex variables $a_\mathbf{k}$, $b_\mathbf{k}$ can be looked at as describing a circularly polarized electromagnetic field in the plane transversal to the direction of propagation $\mathbf{k}$. Assuming field intensity $r$ and polarization angle $\phi$ as conjugate canonical coordinates in this plane, i.e., setting $\displaystyle \mathcal{R}_x$ $\doteq$ $r \cos \phi$, $\displaystyle \mathcal{R}_y$ $\doteq$ $r \sin \phi$, $\displaystyle \mathcal{R}_z$ $\doteq$ $r$, and resorting to the usual $PB$'s, one finds $\{\mathcal{R}_x,\mathcal{R}_y\}_{PB} = \mathcal{R}_z$, $\{\mathcal{R}_z,\mathcal{R}_x\}_{PB} = - \mathcal{R}_y$, $\{\mathcal{R}_z,\mathcal{R}_y \}_{PB} = \mathcal{R}_x$. In other words, the $PB$'s satisfy the algebra $so (1,2)$ $\sim$ $su (1,1)$. Canonical quantization of the $PB$'s should give rise to this Lie algebra: the quantized electromagnetic field amplitude and phase are thus naturally described by $su(1,1)$ rather than $h(1)$. This incidentally avoids the phase--number ambiguity, 
{\sl cf.} \cite{MR,HAKPRA}.
%%%%%%%%%%%%%%%%%%%%%%%%%%%%%%%%%%%%%%%%%%%%%%%%%%%%%%
%%%%%%%%%%%%%%%%%%%%%%%%%%%%%%%%%%%%%%%%%%%%%%%%%%%%%%%%%%
\section{The quantum harmonic oscillator revisited}
%%%%%%%%%%%%%%%%%%%%%%%%%%%%%%%%%%%%%%%%%%%%%%%%%%%
%%%%%%%%%%%%%%%%%%%%%%%%%%%%%%%%%%%%%%%%%%%%%%%%%%%%%%%
The discussion of previous section touches deeply on the nature of dynamical systems described by Hamiltonians bilinear in the ladder operators, unitarily equivalent to the that of the harmonic oscillator (h.o.), assumed as the simplest, fundamental representative of free bosons in all quantum physics (see the thorough review \cite{HAKANN} and references therein). Bilinearity of the Hamiltonian realized in the envelope 
of $h(1)$ straightforwardly reveals (see, {\sl e.g.}, \cite{JAFAROV,KLIMIK}) that the h.o. dynamical algebra is $su(1,1)$. Indeed, \cite{HAKANN,MOSHINSKY}, the algebra in the quadratic envelope of $h(1)$, generated by 
$$
K_1 = \tfrac{1}{4} ( p^2 - q^2 ) \;\;,\;\; K_2 = \tfrac{1}{4} ( q \, p + p \, q ) \;\;,\;\; K_3 = \tfrac{1}{4} ( p^2 + q^2 ) \; , 
$$
with $[q , p] = \imath$, leads to the Schwinger single--boson representation of $su(1,1)$ \cite{SCHWINGER}. Yet position, momentum and Hamiltonian observables  can be defined linearly in the $su(1,1)$ framework but retaining only some of the properties of the $h(1)$ case. A choice which leads to a state space $\sim {\mathfrak{F}}$ is 
$$
\hat{Q} = \tfrac{1}{\sqrt{2}} \, \bigl ( K_+ + K_- )\,;\;\;\hat{P} = \tfrac{\imath}{\sqrt{2}} \,\bigl 
( K_+ - K_- \bigr )\,; \;\; \hat{H} = K_3\; , 
$$
whereby $\bigl [ \hat{H},\hat{Q} \bigr ] = - \imath \hat{P}$, $\bigl [ \hat{H},\hat{P} \bigr ] = \imath \hat{Q}$. These relations show that the $su(1,1)$ variables are compatible with the canonical h.o. Heisenberg equations of motion. The dynamical algebra is manifestly $su(1,1)$, and in ${\cal D}^{(+)}_{\tfrac{1}{2}}$ the eigenvalues of $K_3$ in ${\mathfrak{F}}$ reproduce exactly the energy spectrum of the $h(1)$ quantum oscillator. 
The commutation relation $\displaystyle \bigl [ \hat{Q},\hat{P} \bigr ] = \imath$ is not invoked; here $\bigl [ \hat{Q},\hat{P} \bigr ]$ $=$ $2 \imath K_3$. It is then readily checked that $\hat{H} \neq \frac{1}{2} \bigl ( {\hat{P}}^2 + {\hat{Q}}^2 \bigr ) = K_3^2 - \kappa ( \kappa - 1)$. Different definitions of the relevant operators have been given in this framework \cite{JAFAROV},\cite{KLIMIK}. A consistent realization of position and momentum requires however the full universal envelope of $su(1,1)$. Resorting to the inverse of the Holstein--Primakoff representation \eqref{HP}, $a$ $=$ $(K_3 + \kappa)^{-\tfrac{1}{2}} K_-$, and setting, as usual, $\hat{Q}$ $=$ $(a^\dag + a)/\sqrt{2}$, $\hat{P}$ $=$ $\imath (a^\dag - a)/\sqrt{2}$,  one can check that $\displaystyle [\hat{Q},\hat{P}] = \imath$ and $\hat{H}$ $=$ $\tfrac{1}{2} \bigl ({\hat{P}}^2 + {\hat{Q}}^2 \bigr)$ $=$ $a^\dag a + \tfrac{1}{2} $, with $\hat{H}$ $=$ $K_3 - \kappa + \tfrac{1}{2}$. 
%%%%%%%%%%%%%%%%%%%%%%%%%%%%%%%%%%%%%%%%%%%%%%%%%%%%%%
%%%%%%%%%%%%%%%%%%%%%%%%%%%%%%%%%%%%%%%%%%%%%%%%%%%%%%%
\section{Interacting systems}
%%%%%%%%%%%%%%%%%%%%%%%%%%%%%%%%%%%%%%%%%%%%%%%%%%%%%%
%%%%%%%%%%%%%%%%%%%%%%%%%%%%%%%%%%%%%%%%%%%%%%%%%%%%%%%
In the vast field of interacting boson physics, we focus our attention on the Jaynes--Cummings model \cite{ALLEN} as a test case for our proposal to represent bosons \textit{via} $su(1,1)$ even when bosons interact with matter. In this model the interaction between a two--level ion, characterized by its ground $|g\rangle$ and excited $|e\rangle$ states and transition frequency $\omega_0$ $=$ $\omega_e - \omega_g$, and a quantized single--mode electromagnetic field of frequency $\omega$ is represented by the Hamiltonian 
\begin{equation}
H_{JC} \, =\, \omega \, \hat{n} + \omega_0 S_z + \bigl(\lambda a S_+ + \textrm{H.c.} \bigr ) \; , \label{JCh}
\end{equation}
where the zero--energy level is at the middle of $|g\rangle$ and $|e\rangle$, the light--ion coupling constant $\lambda$ is a $c$-number, and the atomic operators are given by $S_z = \frac{1}{2} \bigl ( |e\rangle\langle e| - |g\rangle\langle g| \bigr )$, $S_+$ $=$ $|e\rangle\langle g|$, $S_-$ $=$ $S_+^{\dagger}$. Since $(\hat{n} + S_z)$ is a constant of motion, Hamiltonian \eqref{JCh} is block--diagonal in the states $|n-1\rangle\otimes|e\rangle$, $|n\rangle\otimes|g\rangle$. For the one-dimensional block $n = 0$ the energy eigenvalue is $- \frac{1}{2} \omega_0$, while for $n$ $\ge$ 1, with $\Delta$ $\doteq$ $\omega-\omega_0$ the detuning, the eigenenergies are 
$$
n \omega -\frac{1}{2} (\Delta + \omega_0) \pm \frac{1}{2} R_n \; ,
$$
with $R_n$ $=$ $\displaystyle \sqrt{\Delta^2 + 4 \lambda^2 n}$ the generalized Rabi frequency $(R_n$ $=$ $2 \lambda \sqrt{n}$ at resonance, $\Delta = 0)$.

The atomic population inversion dynamics, $\langle S_z(t)\rangle$ $\equiv$ $\langle in|S_z(t)|in\rangle$, where $|in\rangle$ is the initial state of the ion-field system, is dealt with in \cite{BUSUK} solving the Heisenberg's equations of motion for $S_z (t)$. Specifically, for $|in\rangle$ $=$ $|g\rangle \otimes |\alpha\rangle$, $|\alpha\rangle$ being Glauber's $h(1)$ coherent states of the radiation field, upon defining, with $\zeta \in \mathbb{C}$,
$$
\mathfrak{S}_{\tau,\mu} (\zeta )\, \doteq \,\sum^{\infty}_{n=0} \,\frac{|\zeta|^{2n}}{(n!)^\mu}\, 
\cos (2 \lambda n^{\tau} t)\,, 
$$ 
the result given in \cite{BUSUK} can be rewritten as 
\begin{equation}
\langle S_z(t)\rangle \,=\, - \tfrac{1}{2} \, \exp (-|\alpha|^2) \, {\mathfrak{S}}_{\tfrac{1}{2},1} \, (\alpha )\; . \label{BSRES}
\end{equation}
Due to the difficulties in the numerical evaluation of ${\mathfrak{S}}_{\frac{1}{2},1}(\alpha )$, an intensity--dependent coupling was proposed, whereby in \eqref{JCh} $\lambda$ is replaced with (operator--valued)  $\lambda_0 \sqrt{\hat{n}+1}$, 
with $\lambda_0$ a $c$-number. This leads in Eq. 
\eqref{BSRES} to the substitution of ${\mathfrak{S}}_{\tfrac{1}{2},1} \, (\alpha )$ with ${\mathfrak{S}}_{1,1} (\alpha )$, readily summable to 
$\exp [\,|\alpha|^2 \cos (2 \lambda_0 t)] \, \cos [\, |\alpha |^2 \sin (2 \lambda_0 t)\, ]$. 
Comparing these two forms for $\langle S_z(t) \rangle$, the time--dependence of the ion flipping from the lower to the upper state and {\sl vice versa} results into a $\sqrt{n}$ or $n$ factor in $R_n$. This induces a significant difference, entailing however only the shortening by a factor $\sqrt{n}$ of the time scale of both the collapse and revival oscillations, physically undetectable.  Another feature of the intensity--dependent model is the well--known phenomenon of periodic decays and revivals of $\langle S_z(t)\rangle$ (see, {\sl e.g.}, the explicit calculations reported in the early works \cite{BUSUK,BUZEK,SINGH}.) Here however we focus on a different aspect of the intensity--dependent model proposed in \cite{BUSUK}, {\sl i.e.}, the fact that the Hamiltonian \eqref{JCh} can be written in terms of $su(1,1)$ generators \cite{BUZEK}
\begin{equation}
H_{JC} = \omega \, (K_3 - \tfrac{1}{2} ) + \omega_0 S_z + \bigl ( \lambda_0 K_- S_+ + \textrm{H.c.} \bigr ) \; . \label{JCsu}
\end{equation}
In \cite{BUZEK}-\cite{RODRIGUEZ}, where various versions of the Buck--Sukumar intensity--dependent model are investigated, this algebraic view is not related to a new representation of photons, whereas here we suggest that the Hamiltonian \eqref{JCsu} is just a further hint to the claim that photons are correctly described in the $su(1,1)$ representation ${\cal D}^{(+)}_{\frac{1}{2}}$ over ${\mathfrak{F}}$, with $K_-$, $K_+$ replacing $a$, $a^\dagger$. Then there is no need of the non--physical hypothesis of a non--linear (intensity--dependent) operator coupling and the conventional picture of a radiation--matter $c$-number interaction coupling constant can be retained. In other words the example of the Jaynes--Cummings model shows that our proposal can be faultlessly implemented for bosons interacting with matter giving a more consistent picture of their behaviour with respect to other approaches reported in literature. Using model \eqref{JCsu} we evaluated $\langle S_z(t)\rangle$ for various initial states and found that it exhibits the same shortening by the factor $\sqrt{n}$ of the periods of collapse and revival. 
 As an example we consider, consistently with Eq.\eqref{JCsu}, $|in\rangle$ $=$ $|g\rangle \otimes |\eta\rangle$, where
$$
|\eta\rangle \,=\, [I_0 (2|\eta|)]^{-\tfrac{1}{2}} \, \sum^{\infty}_{n=0} \frac{\eta^n }{n!} |n\rangle
$$
are the $su(1,1)$ Barut--Girardello coherent states \cite{BARGIR} for $\kappa$ $=$ $1/2$ with $I_0 (z)$  the 0-th order modified Bessel function of the first kind (the case $|in\rangle$ $=$ $|g\rangle \otimes |\xi\rangle$, where $|\xi\rangle$ denotes the $su(1,1)$ Perelomov coherent states \cite{PERE1,PERE2}, is dealt with in \cite{BUZEK}). We 
 have once more a closed--form result, namely
\begin{equation}
\langle S_z(t)\rangle \,=\, - \tfrac{1}{2} \, [ I_0 (2|\eta|)]^{-1} \, {\mathfrak{S}}_{1,2}( \eta ) \; , \label{SZBG}
\end{equation}
$\mathfrak{S}_{1,2}(\eta)$ $=$ ${\rm{Re}} \bigl ( I_0 (2|\eta| e^{\imath \lambda_0 t}) \bigr )$. Using instead Hamiltonian \eqref{JCh} would entail replacing in Eq. \eqref{SZBG} $\mathfrak{S}_{1,2}(\eta)$ with ${\mathfrak{S}}_{\tfrac{1}{2},2}(\eta)$ which is once more a slowly convergent series that can be evaluated only numerically, not without difficulties.
%%%%%%%%%%%%%%%%%%%%%%%%%%%%%%%%%%%%%%%%%%%%%%%%%%%%%%
%%%%%%%%%%%%%%%%%%%%%%%%%%%%%%%%%%%%%%%%%%%%%%%%%%%%%%%
\section{Conclusions}
%%%%%%%%%%%%%%%%%%%%%%%%%%%%%%%%%%%%%%%%%%%%%%%%%%%%%%
%%%%%%%%%%%%%%%%%%%%%%%%%%%%%%%%%%%%%%%%%%%%%%%%%%%%%%%
We have presented a series of theoretical arguments to sustain the claim that resorting to the algebra $su(1,1)$ provides a more effective, more consistent, more correct scheme to describe bosons than the conventional $h(1)$. This statement is grounded on the statistical counting for bosons 
(quantum {\sl vs.} classical) and on the $su(1,1)$ invariance of the Maxwell equations (relativistic {\sl vs.} classical). The question has a strong weight as a matter of principles. The impact in terms of corrections to phenomenological features, especially for interacting boson systems described  within
 the ensuing unified scheme, needs to be systematically explored.
%%%%%%%%%%%%%%%%%%%%%%%%%%%%%%%%%%%%%%%%%%%%%%%%%%%%%%%
%%%%%%%%%%%%%%%%%%%%%%%%%%%%%%%%%%%%%%%%%%%%%%%%%%%%%%%
\ack
AM acknowledges partial support from PRIN 2010-2011 $'$Geometrical and analytical theories of finite and infinite dimensional Hamiltonian systems$'$.\\
FAR and  MR acknowledge the financial support of Compagnia di San Paolo (Torino, Italy) in the frame of the INRiM project on $'$Quantum Correlations$'$.
%%%%%%%%%%%%%%%%%%%%%%%%%%%%%%%%%%%%%%%%%%%%%%%%%%%%%%%
%%%%%%%%%%%%%%%%%%%%%%%%%%%%%%%%%%%%%%%%%%%%%%%%%%%%%%%

\end{document}